\documentclass[prb,aps,twocolumn,superscriptaddress]{revtex4}

\usepackage[usenames,dvipsnames]{xcolor}

\newcommand{\hi}{\normalcolor}
\newcommand{\hiB}{\normalcolor}
\newcommand{\lo}{\normalcolor}

\usepackage[pdftex]{graphicx}

\usepackage{verbatim}
\usepackage{amsmath}
\usepackage{amsfonts}
\usepackage{amssymb}
\usepackage{fancyvrb}

\newcommand{\scgw}{\text{sc}GW}
\newcommand{\io}{IO} 
\newcommand{\tpp}{t^{\prime\prime}}
\newcommand{\Uo}{U}

\newcommand{\NR}{N_{\mathrm{R}}}
\newcommand{\kD}{k_{\mathrm{D}}}

\newcommand{\HH}{{\cal{H}}}

\begin{document}
\title{Incommensurate quantum-size oscillations in acene-based molecular wires 
- effects of quantum fluctuations} 
\author{Peter \surname{Schmitteckert}}
\email{peter.schmitteckert@physik.uni-wuerzburg.de}
\affiliation{Institute for Theoretical Physics and Astrophysics, Julius-Maximilians University 
of W\"urzburg, Am Hubland 97074 W\"urzburg, Germany}
\author{Ronny \surname{Thomale}}
\affiliation{Institute for Theoretical Physics and Astrophysics, Julius-Maximilians University 
of W\"urzburg, Am Hubland 97074 W\"urzburg, Germany}
\author{Richard \surname{Koryt\'ar}}
\affiliation{Institute of Theoretical Physics, University of Regensburg, 93040 Regensburg, Germany}
\affiliation{Department of Condensed Matter Physics, Faculty of Mathematics and Physics, Charles University in Prague, 121 16 Praha 2, Czech Republic}
\author{Ferdinand Evers}
\affiliation{Institute of Theoretical Physics, University of Regensburg, 93040 Regensburg, Germany}
\date{\today}

\begin{abstract}
Molecular wires of the acene-family can be viewed as 
a physical realization of a two-rung ladder Hamiltonian. 
For acene-ladders, closed-shell ab-initio calculations and 
elementary zone-folding arguments predict incommensurate 
gap oscillations as a function of the number of repetitive ring units, $N_{\text{R}}$, 
exhibiting a period of about ten rings. 
Results employing open-shell calculations and a 
mean-field treatment of interactions
suggest anti-ferromagnetic correlations that could 
potentially open a large gap and wash out the gap oscillations.
Within the framework of a Hubbard model 
with repulsive on-site interaction, $U$, we employ a Hartree-Fock
analysis and the density matrix renormalization group to investigate the
interplay of gap oscillations and interactions.
We confirm the persistence of incommensurate 
oscillations in acene-type ladder systems for a significant fraction
of parameter space spanned by $U$ and $N_{\text{R}}$. 

\end{abstract}
\maketitle

\section{Introduction} 

The question how properties of a macroscopic 
system emerge when more and more atoms or molecules 
accumulate exhibits many different 
facets and, for that reason, reappears every once in a while 
in different contexts. 
On the simplest level of ``emergence'', 
one could consider the appearance of band-structures
in crystal growth. 
A first qualitative description  
of this phenomenon can already 
be given within a picture 
of non-interacting quasi-particles. 
A conceptually analogous situation arises in mesoscopic 
physics: when a quantum dot couples to an electrode,
the dot states hybridize with contact states, 
thereby acquiring the level broadenings, $\Gamma$, 
that signal finite lifetime effects.
Qualitatively new physics emerges 
upon accounting for interaction effects. 
For instance, 
in quantum-dots, the Kondo phenomenon results from the 
interplay of hybridization and the Coulomb-energy. 
Another example displaying the appearance of 
strong correlation effects with growing system size 
is superconductivity. 
\cite{vonDelft,EfetovReview}
There, one is facing a crossover scenario, 
where the super-conducting gap, $\Delta_{\text{sc}}$, 
competes with the single particle level spacing, 
$\Delta_0\gg \Gamma$. 

In this paper, we investigate an incarnation 
of a similar motif as it appears in {\em molecular electronics}: 
we ask how the electronic spectral 
properties of a molecular wire built out of $N_{\text{R}}$ 
repetitive units (``rings'') evolve with increasing $N_{\text{R}}$. 
Motivated by earlier experiments 
\cite{kiguchi2008highly,yelin2013,yelin2015},
our specific example will be the acene-family such as benzene,
naphthalene, anthracene, and others. 
Due to their peculiar electronic structure, oligo-acenes have
received a considerable amount of attention in 
organic chemistry over the past 
50 years.\cite{bettinger10,bettinger16,yang16}
A recent technological interest is based on suggestions 
to use them as light harvesters in organic 
photo-voltaic devices.
\cite{dabo13}
Besides being a simple realization of nano-wires, 
acenes can be viewed as the narrowest graphene nano-ribbons with
zig-zag terminated edges. 
Due to recent progress in synthesis methodology, 
oligoacenes up to length $N_{\text{R}}=9$ (nonacene) 
have already been produced.\cite{toenshoff10,bettinger15} 

\begin{figure*}[t]
  \begin{minipage}{0.45\textwidth}
   \includegraphics[width=0.8\columnwidth]{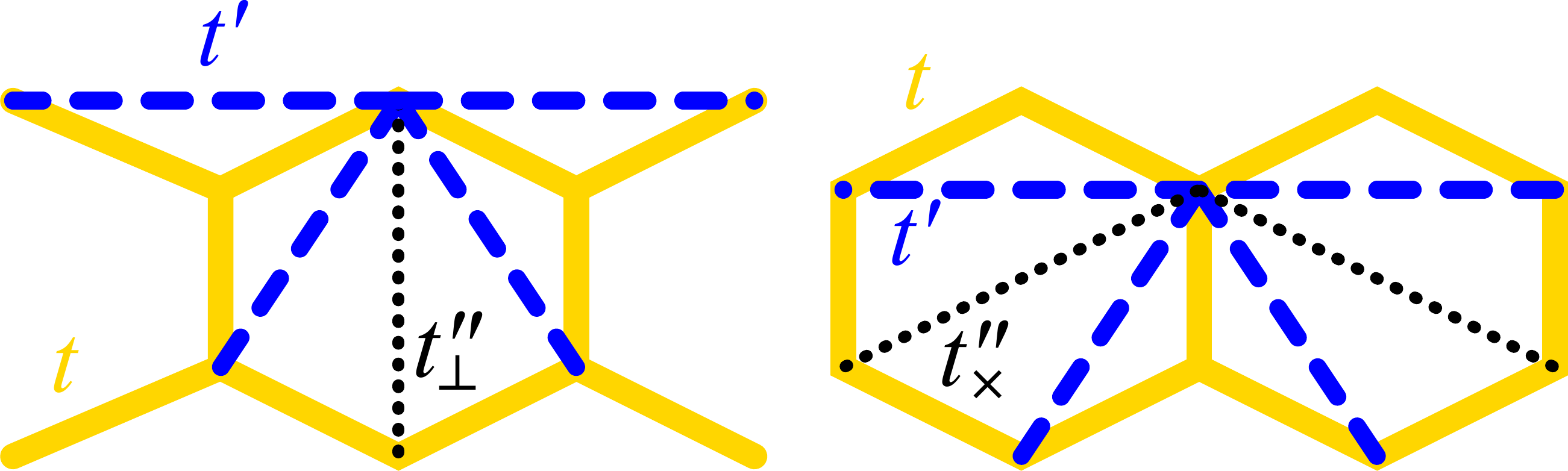}
  \end{minipage}
 \begin{minipage}{0.45\textwidth}
   \includegraphics[width=0.8\columnwidth]{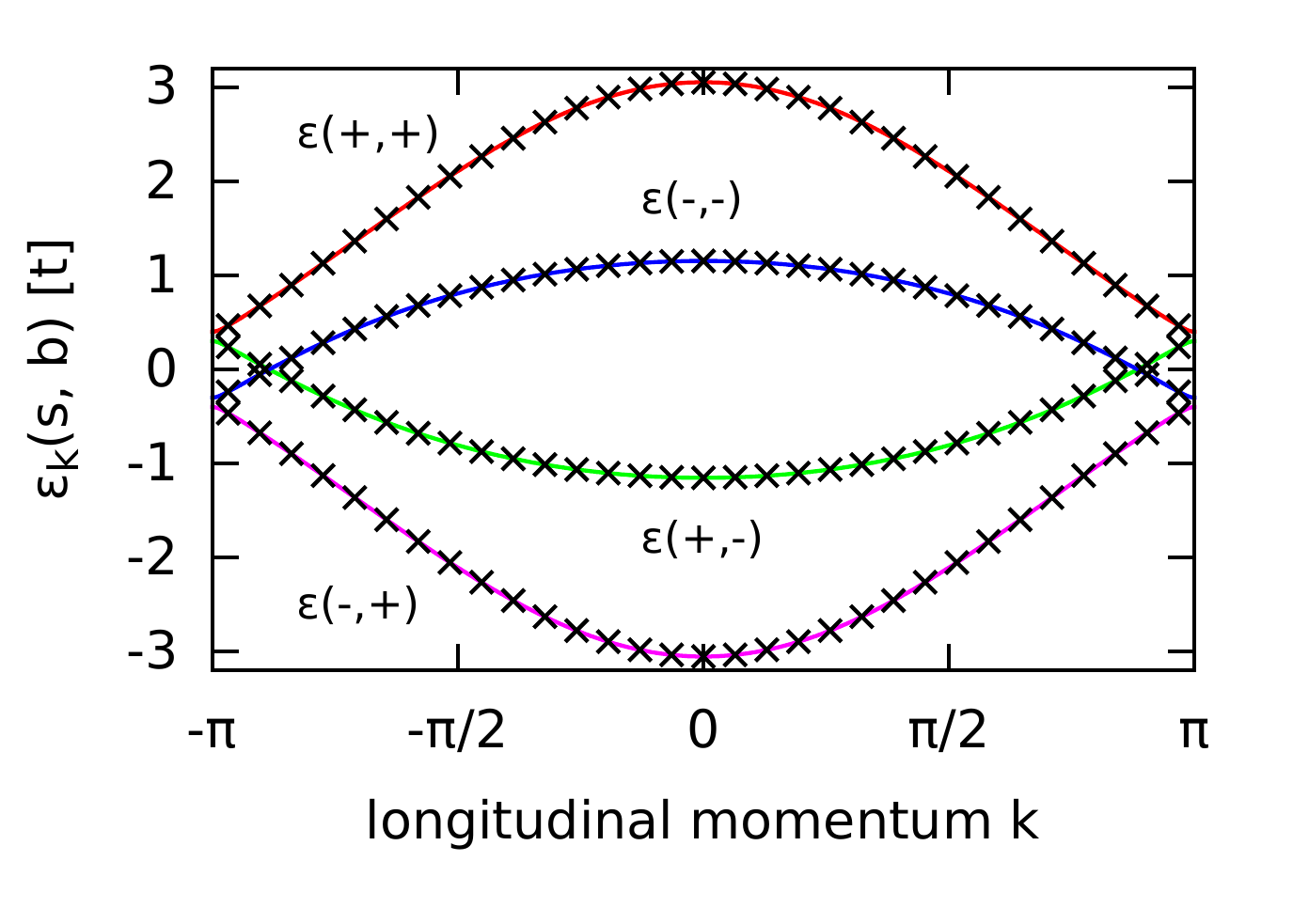}
  \end{minipage}
\caption{\label{fig:lattice}
  (Left) Parts of the oligoacene backbone in the tight-binding representation
  with a nearest neighbor hopping ($t$, yellow),
  hoppings to the second neighbor ($t^\prime$, dashed blue)
  and third neighbor ($\tpp$, dotted black). On the left, we draw these
  hoppings for the "outer" carbon site, on the right for the "inner" carbon.
  $\tpp_\perp$ denotes the second neighbor hooping of the outer sites, $\tpp_\times$ of
  the inner sites.
  In the present study we adopt $\tpp=\tpp_\perp=\tpp_\times$ and, in contrast to Ref. \cite{korytar2014},  
   $t^\prime{=}0$
  \\
  (Right) The  tight binding band-structure corresponding
  to the long-wire limit for $t=1$, $t'=0$, and $\tpp=0.3$. 
  The lines are the analytical result according to \eqref{disp} with  $\tpp_\perp = \tpp_\times$.
  The crosses are the results from a tight binding system consisting of $\NR=31$ rings and periodic boundary conditions.
  The valence- and conduction-band are crossing  at the $\Gamma-$point $\kD$. 
  }
\end{figure*}

Previously, we have found that linear
(oligo)acenes exhibit oscillations in the $N_{\text{R}}$ (or length) dependence of
optical excitation gaps. \cite{korytar2014}
Oscillations of electronic properties with the geometric size of a
nano-system are {\it per se} frequently encountered. 
An established example is given by the threefold gap-oscillation 
of zigzag carbon nanotubes (armchair-ribbons) as a function of tube diameter (width). 
\cite{reichBook,brey06,dasgupta12} 
In this context, the three-fold periodicity reflects the
symmetry of the hexagonal Brillouin zone, with the 
Dirac points located at the corners.  
A remarkable aspect of the acene oscillations is that they 
are incommensurate in the sense that the oscillation 
period is not dictated by lattice symmetry; it can reach 
periods of ten times the length of the unit cell. 
The oscillations come with an important consequence,
namely, a close-to closing gap for molecules of certain 
(periodically repeating) finite length. 

We have demonstrated 
that similar to the case of the nanotubes, 
the acene-oscillations can likewise be described 
(nearly) quantitatively by a selection rule 
({\em zone-folding}) for the wave-vectors 
applied to the band structure of the underlying 
extended tight-binding model (Fig.~\ref{fig:lattice}).
The procedure is analogous to the threefold periodicity in band-gaps of
nanotubes\cite{reichBook} and nanoribbons\cite{brey06,dasgupta12}; 
in the acene case, however, incommensurate periodicity is reached because of 
the non-universal position of $k_\text{D}$ labelling the longitudinal
momentum at the acene-analogue of the Dirac point.
Based on the computational analysis of the acene-like Hubbard model,
we find that the zone-folding argument is not invalidated 
by interactions beyond the independent quasi-particle picture; 
the oscillations persist given the interaction strength $U$ 
does not exceed a threshold of the order of the bandwidth - at least
as long as interactions are screened, i.e., short-ranged.

\par
Building upon our previous work, we intend to give a comprehensive
study of interactions and incommensurate oscillations (IO) of acenes
by 
contrasting Hartree-Fock (HF) mean field treatments against microscopic
numerical studies. We model acene-type molecular wires by
finite length Hubbard ladders. 
A principal physical incentive is given by the
potentially rich phase diagram of interacting acene wires
\cite{karakonstantakis13}. In this context, one could hope to
manipulate acenes via chemical synthesis,  or engineering of the
environment so as to explore some part of this phase diagram. 
\hiB
(Admittedly before this is experimentally feasible, major 
technological difficulties have to be overcome that are realated 
to chemical stability. Momentarily, 
long acene wires need to be kept within a noble-gas matrix 
to prevent further reactions\cite{bettinger15}.)
\normalcolor

Further methodological incentive is provided by recent 
open-shell $\scgw$-calculations for the acene-series\cite{wilhelm16}; 
they indicated that in short wires the \io{} might {\em disappear} 
in the presence of anti-ferromagnetic {\em order} in the ground state. 
At first sight, this result may look very plausible. 
However, effects of quantum-fluctuations are neglected 
in its derivation and in quasi-onedimensional systems 
their effect typically is very strong. 
Specifically, in the long-wire limit, 
quantum fluctuations tend to destabilize 
mean-field ordered phases that break a continuous symmetry~\cite{PhysRevLett.17.1133}.
In this (quantum-disordered) phase, 
the order-parameter exhibits
significant short-range correlations
that leave a characteristic signature in local probe 
measurements.

In this work, we clarify the fate of the anti-ferromagnetic 
correlations in acenes-type ladders in the presence of quantum 
fluctuations, i.e., by providing a joint analysis from Hartree-Fock (HF)
mean field and density matrix renormalization group (DMRG). 
Within HF mean field analysis, we reproduce the observation 
made in the ab-initio study Ref. \onlinecite{wilhelm16} for 
acene-molecules with Coulomb interactions: 
acene-ladders with onsite Hubbard interactions exhibit anti-ferromagnetic order on the mean field level along with a
large charge gap, thereby diminishing \io{}.
By microscopic DMRG analysis, however,
we show that also in short ladders quantum fluctuations 
destroy the long-range order. 
In particular, they significantly reduce the HF 
excitation gap, restoring the \io{} signature and
up to intermediate interaction strengths
of the order of the bandwidth.


\section{Model and Methods}
\subsection{Model definition} 
We study the acene Hubbard-ladder
\begin{equation}
\newcommand{\half}{\frac{1}{2}}
\label{e1}
\hat H =
\sum_{i\neq j}t_{ij} \hat c^\dagger_{i\sigma}\hat c^{\phantom\dagger}_{j\sigma}
+ \Uo\sum_i \left ( \hat n_{i\uparrow} - \half\right )
\left (\hat n_{i\downarrow} - \half \right )
\end{equation}
at half filling, where the first term describes the tight-binding
dynamics, and  the second term in Eq. \eqref{e1} the onsite Hubbard repulsion with strength $\Uo$. 
In the case of periodic boundary conditions (PBC), we have a linear length of $L= 2\NR$ sites,
where $\NR$ gives the number of hexagon (benzene) rings.
For hard wall boundary conditions (HWBC), we complete the last ring leading to $L= \NR + 1$.
The total number of sites is $2L$.
The nearest neighbor hopping is characterized by an 
amplitude $t_{\langle ij\rangle}=-t$, and the longer-range hoppings
accordingly by $t'$, $\tpp_{\perp}$, and $\tpp_{\times}$ (Fig.~\ref{fig:lattice}). 
If not stated otherwise, $t$ will be taken as the unit of energy
and $\tpp = \tpp_{\perp}$ = $\tpp_{\times}$.
Note that $\tpp_{\perp}$ and $\tpp_{\times}$ give rise to a non-avoided
crossing of conduction- and valence  band at non-commensurate
wave-vectors $k_\text{D}$ predicted  for the case
$\Uo{=}0$.\cite{Kivelson} For $\tpp_{\perp} ,\tpp_{\times}=0$, the system features a
quadratic band touching~\cite{karakonstantakis13}.
For $U=0$, the energy spectrum of~\eqref{e1} reduces to 
\begin{widetext}
\begin{equation} \epsilon_{sb}(k) = 2t'\cos(k) + \frac 12 s\left(t + \tpp_\perp + 2 \tpp_\times \cos(k)\right) 
  + \frac{sb}{2} \sqrt{\left(t - \tpp_\perp + 2\tpp_\times  \cos(k)\right)^2 + 8(1+\cos(k))(t + st')^2}, \label{disp}
\end{equation}
\end{widetext}
where $b,s=\pm 1$ denote a band and parity index, respectively,
yielding four bands resulting from the four-site unit cell. 

In the following, we exclude next-nearest
neighbor hopping, $t^\prime=0$, so there is no hopping  
between sites sharing the same sublattice; the model becomes 
bipartite and a chiral symmetry is imposed connected to the
particle-hole symmetry of the spectrum (Fig.~\ref{fig:lattice}).
This choice can be made without loss of generality regarding \io{}, as the main effect of breaking the chiral symmetry 
in more general models with non-vanishing $t^\prime$ would be to shift the position of
the Dirac point $k_\text{D}$. 
The corresponding analytical calculations are given in
the Appendix~\ref{sec:appendixKorytar}.

\hiB
{\em General comments on Hubbard-models.} 
We briefly comment about the applicability of Hubbard models to real
systems. Indeed, models along the lines of \eqref{e1} 
have been investigated intensively over the last fifty years in
physics and in chemistry. 
 They provide substantial simplifications by (a) ignoring the coupling
 to molecular vibrations; (b) including only one orbital per atom and
 (c) by including all other degrees of freedom in terms of effective
 parameters, only, such as a short range interaction that mimics
 screening. In fact, information about the molecular geometry
(e.g. the relative postion of the atoms) is absorbed in hopping
parameters and interaction matrix elements. These parameters span 
a multi-dimensional space which has a hyperplane that features 
those particular combination of parameters that can be 
reached in realistic situations.

In return, because Hubbard models are in a certain sense minimal, 
they allow for a numerically exact treatment of the many-body problem,
with a size of the computational Hilbert space that
 is much larger as compared to most other many-body
treatments that attempt to be more ``first principles''  
on the Hamiltonian  level. 

Keeping this in mind, Hubbard models like \eqref{e1} are by no means as
oversimplistic as they may appear at first sight. First, if properly parametrized 
such models have a good chance to give a realistic, sometimes even quantative
description of low-energy excitations also in those situations where quantum-fluctuations are
dominant, so that most competing methods fail. 
Second, one often is interested in the system behavior under varying
external conditions, i.e. in the phase diagram of the model; 
one would like to figure out what phases exist, what are their
properties  and, ideally, also where are  the 
phase-boundaries located at. 
Since the qualitative aspects of phase diagrams tend to be very robust under 
a change of the Hamiltonian,  one often adopts convenient choices for the model
parametrization, even though they are sometimes far from realistic. 

Once the phase-diagram has been mapped out, 
it may be possible to predict the qualitative behavior for a given 
experiment, even though the actual parameters representative for 
this situation may not be known with  high accuracy, at least in those 
situations where very stable phases have been identified. 
\normalcolor

\subsection{Computational details}

In this spirit, we continue our analysis and treat the model Hamiltonian \eqref{e1} 
numerically within DMRG~\cite{White:PRL92,White:PRB93}.
%
Similar to previous authors\cite{Jiang, Chan},  
we also observe that the ground state is a spin singlet independent of the 
wire length. Thereby, we track the many-body Hamiltonian at 
fixed particle number (equal to the number of lattice sites)
and in the $S^z_{\text{tot}}{=}0$ spin sector.
By targeting not only the ground state but also excited states, we obtain the optical gap 
by taking the energy difference between the first excited and the ground state.
For non-interacting systems, the optical gap corresponds to the HoMO/LuMO gap, as it is given
by the lowest particle-hole excitation.
Due to spin-rotational invariance, the excited states 
of the model in~\eqref{e1} come in spin-multiplets according to
irreducible representations of SU(2). 
By performing reference calculations 
in the spin-sectors $S^z_{\text{tot}}{=}1,2$, we have ascertained that the lowest 
excitation is a triplet state, and that the ground state is indeed a singlet state.
Details of the DMRG implementation are given in Appendix~\ref{sec:HFNumericsDetails}.
To investigate the effect of quantum fluctuations, and to link with
previous mean field treatments, 
it is useful to compare the DMRG results to a HF
description.  We have also implemented a self-consistent HF
cycle to obtain mean field estimates for the energy gaps, ground state, and
magnetization (Appendix~\ref{sec:HFNumericsDetails}). In HF theory, we look for the lowest lying (charge-neutral) excitation 
energy to determine the energy gap, for which we compute the
difference in energy between the lowest 
unoccupied and the highest occupied 
eigenvalue of the Fock operator. 

\section{Results and Discussion} 

\subsection{Hartree-Fock analysis} 

\begin{figure}[tb]
  \includegraphics[width=0.9\columnwidth]{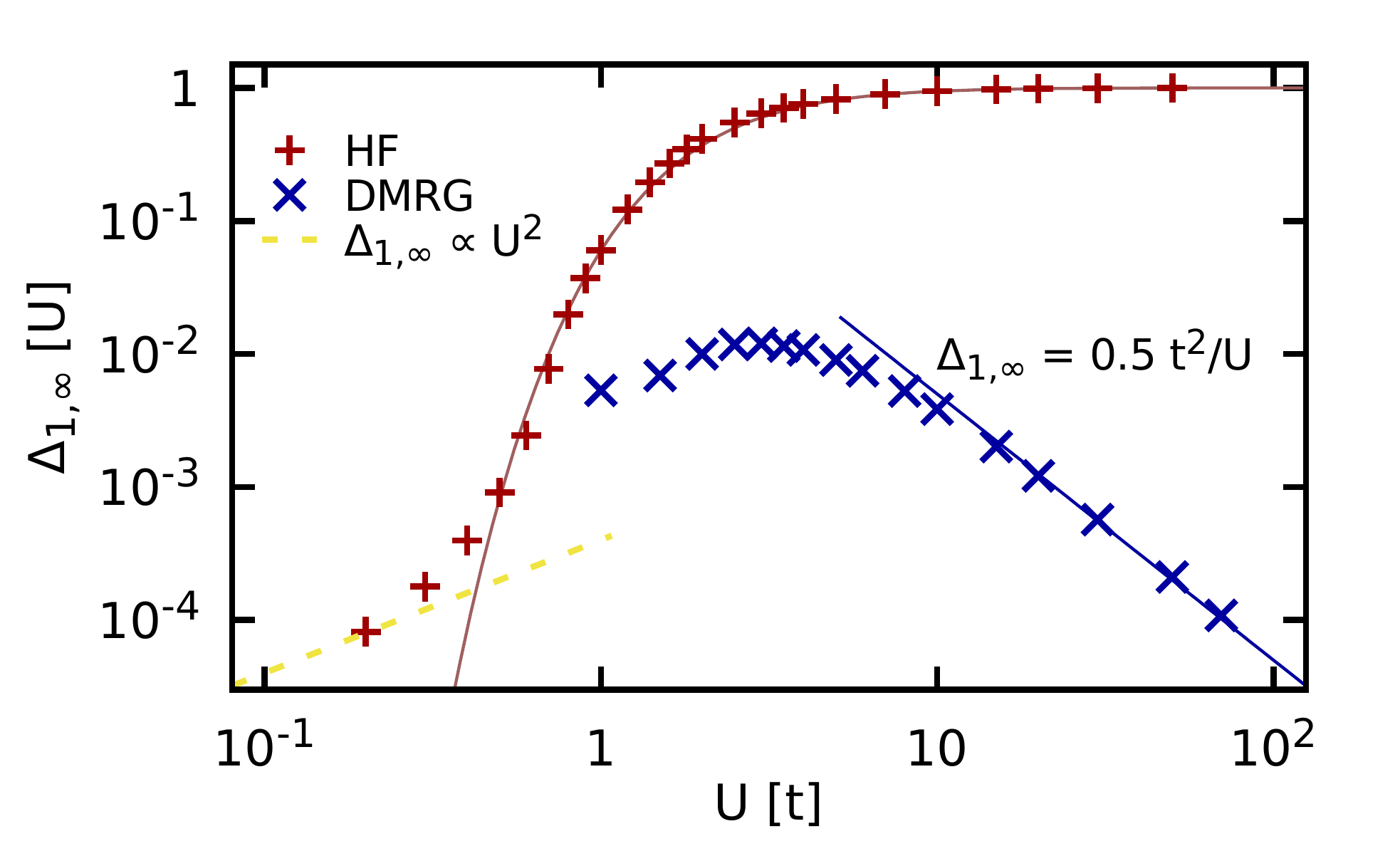}\\
   \includegraphics[width=0.9\columnwidth]{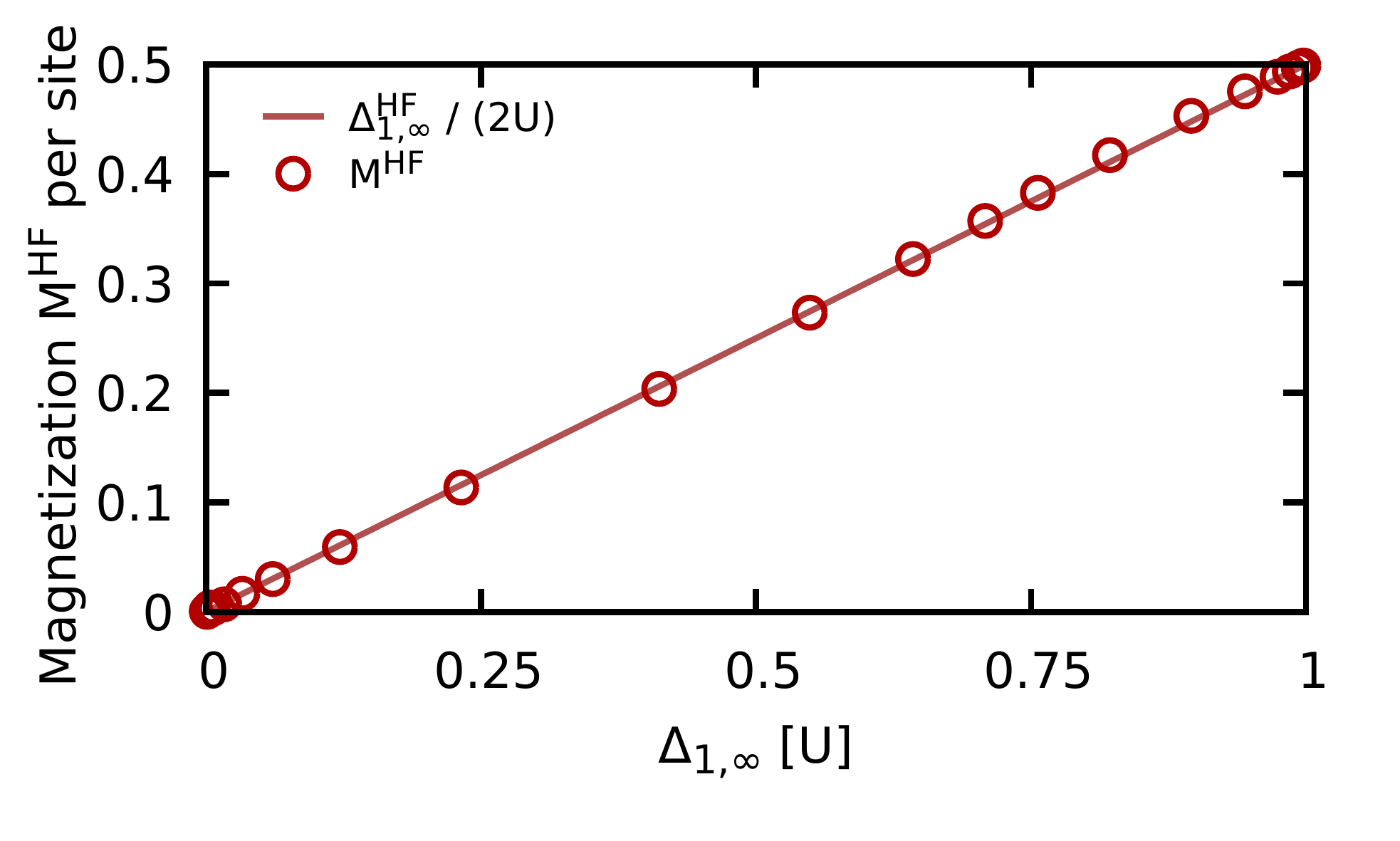}
  \caption{\label{f3a}\label{fig:GapsIO}
  Top: Evolution of the excitation gaps $\Delta^\text{HF}_{1,\infty}(\Uo)$
  ($+$) and $\Delta^\text{DMRG}_{1,\infty}(\Uo)$ ($\times$) 
  in units of the interaction $\Uo$ vs.\ the increasing strength of the repulsive Hubbard interaction $\Uo$.
  For large $\Uo$ we recover the expected asymptotic behavior.
  For $\Uo \rightarrow 0$ the numerical extrapolation is impaired by 
  the \io{}; the asymptotic behavior is indicated by the (yellow) dashed 
  line; see text and for comparison Fig.~\ref{fig:HFgapCO}.\\
  Bottom: Corresponding staggered magnetization $M^\text{HF}$ ($\circ$)
  over the band-gap $\Delta^\text{HF}_{1,\infty}(\Uo)$ in units of $\Uo$ displaying a linear dependence in the complete
  regime.
 }
\end{figure}

Before investigating finite size features of the spectrum, we
analyze the infinite wire limit.
In Fig. \ref{f3a} (top), we display the evolution of the 
HF gap for repulsive interaction strength $\Uo$.
We extract the bulk gap from finite-size calculations by 
extrapolating the finite size gap vs. the inverse linear 
length of the system (Appendix~\ref{app-gap}). 
While the non-interacting wire exhibits a band crossing 
at the Fermi-energy and therefore is metallic, 
the repulsive interaction $\Uo$ 
opens up a correlation gap. 

Within HF theory, the ground state is found to be unstable against
antiferromagnetic order; the corresponding staggered magnetization 
is shown in Fig. \ref{f3a} (bottom). 
As we will see in our DMRG-calculations below, 
this order will be destroyed by quantum fluctuations. 
We here witness the established fact that
mean-field theories overestimate the tendency for symmetry breaking, 
at least if the interactions are short ranged. 
In passing, we mention that broken symmetries refer to the long-distance properties 
of a quantum state, which are not well described on the HF-level. 
In contrast, the total energy is predominantly dictated by the 
pair-correlation function at short-distances. The latter is largely 
insensitive to the global symmetries. Therefore, with respect to 
total energy calculations, HF can often give useful answers even 
though the long-range physics is described qualitatively incorrect. 
(L\"owdin's ``symmetry breaking dilemma'' of HF\cite{lykos63}) 

In the perturbative limit of weak repulsion, $U {\ll} 1$, 
we can restrict ourselves to states with AF ordering on top of
the non-interacting ground state. In this situation 
the gap opens in long (but finite-length) wires as
\begin{equation}
  \Delta^{\mathrm{HF}}_{1,\NR} = \sqrt{ \Delta_0^2 + (M^{\mathrm{HF}} U)^2 } \,,
\end{equation}
where $\Delta_0$ is the finite size gap 
of the non-interacting wire, $\Delta^{}_0 \equiv \Delta^{U=0}_{1,\NR}$; 
$ M_{\mathrm{HF}}$ denotes the staggered
magnetization amplitude. We recall that at $U{\ll}1$
the response of energies to the staggered
field is linear in $U$, implying a quadratic gap opening
$\Delta^\text{HF}_{1,\infty}(\Uo)\sim U^2$
in the thermodynamic limit 
(where $\Delta_0 \rightarrow 0$).

At intermediate interactions, the HF values for the gap opening 
allow for a phenomenological two-parameter fit, 
\begin{equation}
\Delta^\text{HF}_{1,\infty}(\Uo)\approx 0.5\Uo e^{\frac{-D}{\Uo^2 +D_1
    \Uo}} ,
\end{equation}
where $D$ is of the order of the non-interacting bandwidth; we obtain $D\approx 5.5\pm0.4$ and 
$D_1\approx 1.0\pm 0.1$, provided $\Uo$ is not too small. As expected,
in the large $U$ limit, the HF particle-hole gap is set by $U$.
Its evolution follows closely the one of the anti-ferromagnetic 
order parameter $M^\text{HF}(\Uo)$, Fig.~\ref{f3a} (bottom).

We now turn to the HF analysis of acene wires with finite length. 
In this case, one expects that the opening of the HF gap 
$\Delta^\text{HF}_{1,N_{\text{R}}}$ 
 decreases the ground-state energy only 
if $\Delta^\text{HF}_{1,N_{\text{R}}}$  
exceeds the non-interacting 
(``native'') level-spacing $\Delta_0$ 
at the Fermi energy significantly. 
Hence, for shorter wires, the 
gap opening is most effective
at particular values of $\NR$
that have an anomalously small single-particle gap
-- and ineffective at the others. In other words, only at 
$\NR$-values situated close to 
a minimum of the \io{} an AF-order develops and at the 
others the non-interacting ground-state prevails. 
In this situation 
\io{} remain visible even in $\Delta^\text{HF}_{1,\NR}$.

\begin{figure}[tb]
  \includegraphics[width=0.9\columnwidth,type=pdf,ext=.pdf,read=.pdf]{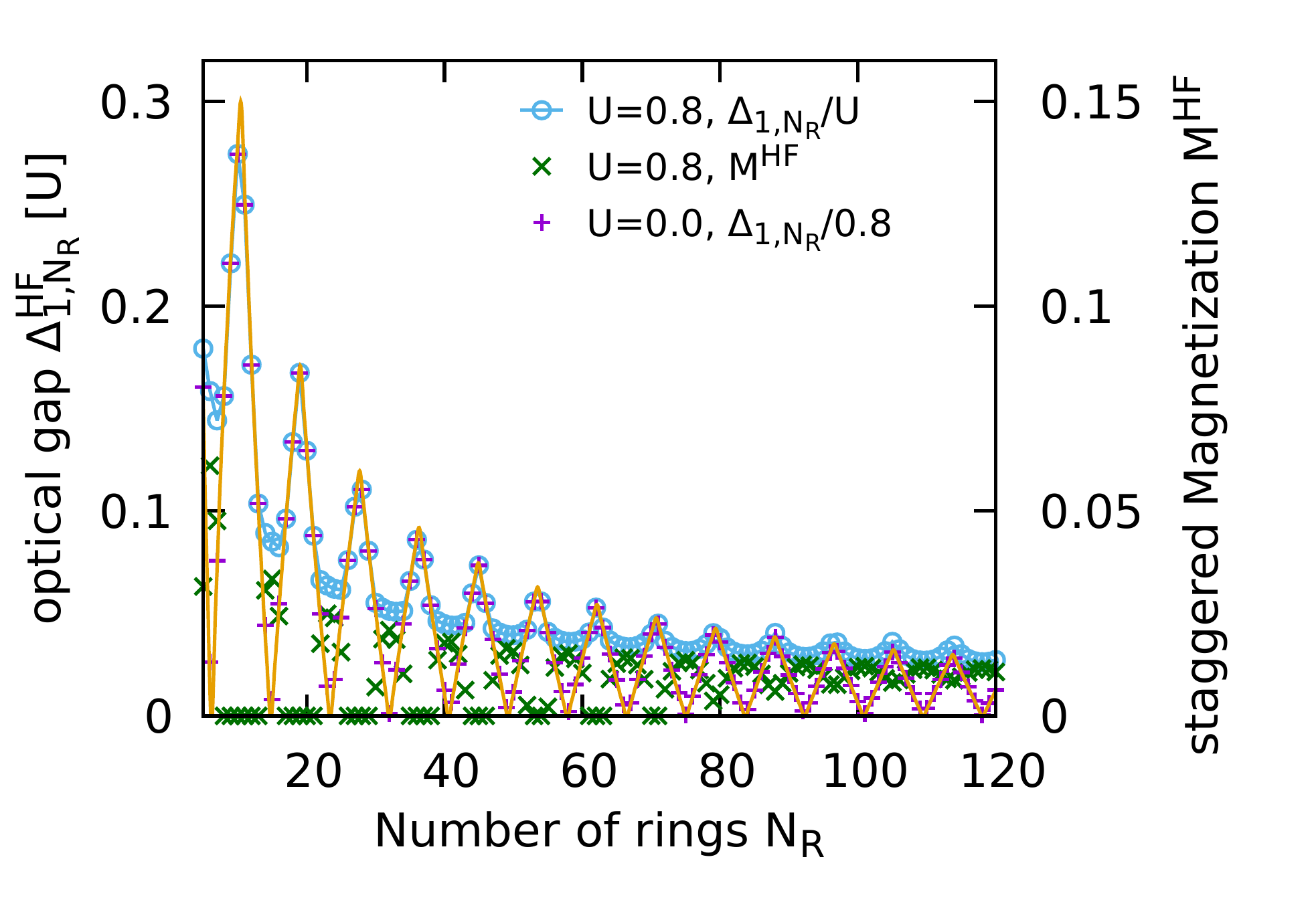}
  \caption{\label{f4}
  Evolution of the HF-magnetization $M^{\mathrm{HF}}$ and HF-gap $\Delta^{\mathrm{HF}}_{1,\NR}$ with 
  increasing number of rings $\NR$ at $\Uo{=}0.8$ and HWBC.
  As is seen, the HF magnetization $M^{\mathrm{HF}}$ of 
  the acene-wire ($\times$) 
  is non-vanishing whenever the HF-gap ($\circ$) exceeds 
  the non-interacting gap ($+$). 
  For this reason, even the mean-field magnetization exhibits  
  the \io{}, in principle. 
  They fade away only for large system sizes, 
  when the non-interacting finite site gap falls below the interaction induced HF gap.
}
\end{figure}

For longer wires, however, the \io{} fade away, 
because the native band-gap is suppressed 
as $\sim 1/\NR$ and eventually no longer overcomes $\Delta^\text{HF}_{1,N_{\text{R}}}$. 
As soon as $\Delta^\text{HF}_{1,N_{\text{R}}}$  exceeds 
the maxima of the native band gaps, the 
\io{} disappear from the HF spectra. 
Indeed,
the \io{} in $\Delta^\text{HF}_{1,N_{\text{R}}}$ 
are seen to 
become weaker with increasing $\NR$ 
(and $\Uo$) (Fig.~\ref{f4}). 
This explains why large systems sizes $\NR {\gg} 1$ are needed 
in order to obtain the HF gap in the small $U$ regime: 
the finite size induced gap has to fall below the HF gap.

As the magnetization closely follows $\Delta^\text{HF}_{1,\NR}$,
it emerges preferably in
acene wires that exhibit a relatively 
small (native) band-gap (Fig.~\ref{f4}). 
Vice versa,    
the HF-ground state remains non-magnetized  at  
wire-lengths, such as $\NR=11,19,28,\ldots$, 
where the native gap is particularly large; there mean-field magnetism
only appears at 
larger interaction values. 
\begin{figure}[b]
  \includegraphics[width=0.9\columnwidth,type=pdf,ext=.pdf,read=.pdf]{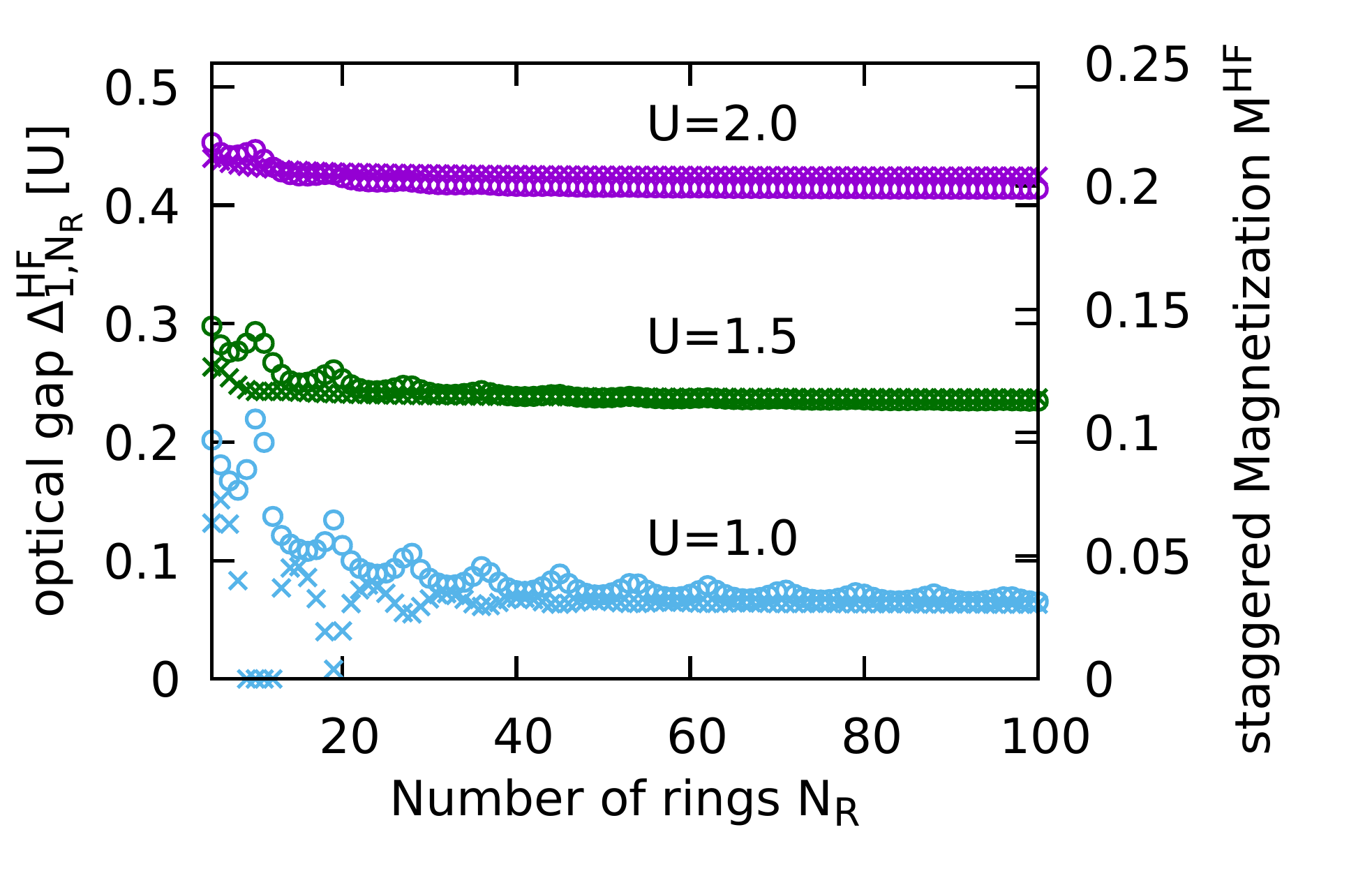}
  \caption{\label{f4b}
  Evolution of the HF-magnetization ($\times$) and HF-gap $\Delta^{\mathrm{HF}}_{1,\NR}$ ($\circ$) with 
  increasing number of rings $\NR$, $L=2\NR+1$, at $\Uo{=}1.0$, $1.5$, $2.0$.
  As in Fig.~\ref{f4} the \io{} vanish if the HF gap is larger than the non-interacting gap. 
  Accordingly, at these instances the staggered magnetization $M^{\mathrm{HF}}$ stays finite.
}
\end{figure}
In contrast, for large interaction values the HF gap surpasses the non-interacting gap
already for small system size (Fig.~\ref{f4b}), and the \io{} disappear in that regime.

\subsection{DMRG calculations} 

We reignite the discussion from the viewpoint of DMRG by the
infinite wire limit. Within our DMRG-implementation, this limit can only be achieved through finite size extrapolation: 
for every value of $U$, the extrapolation is performed to yield
one data point of the blue curve for the excitation gap in
Fig. \ref{f3a} (top), where it is compared against the HF result. 
We mention that at small values of $U$ numerical calculations 
are more challenging to converge in DMRG, so that the 
low $\Uo$ asymptotics is not fully resolved in Fig. \ref{f3a}(top). 
The difficulty is that due to the metallic character of the wire, 
the entanglement content is spread over a large manifold of modes that
have non-negligible weight in the reduced density matrix.

By comparing to the HF result, one directly observes the effect of strong   
quantum fluctuations in the DMRG-trace shown in Fig. \ref{f3a} (top). Fluctuation effects are visible by a reduction of
the excitation gap,
but also by a modified gap evolution with $U$ as compared
to the HF trajectory.
The asymptotic power-law  at 
large $\Uo$ seen in Fig.~\ref{f3a}
is readily understood in terms of the mapping to the 
Heisenberg spin-chain. It proceeds via the effective coupling constant 
$J(\Uo) = 4 t^2/U$ that sets the scale for 
$\Delta^\text{DMRG}_{1,\infty}$ in this limit. \cite*{Anderson:PR50,Anderson:PR59,Bulaevskii:ZETF66,Takahashi:JPC77}
As is seen in Fig. \ref{f3a} (top), the Heisenberg limit is reached 
at values of $\Uo$ exceeding the bandwidth, $D$, 
by about a factor of two. 
\footnote{We adopt the ratio 
$\Delta^\text{HF}_\infty/\Delta^\text{DMRG}_\infty$
taken at a cross-over scale, $D$, 
to quantify the strength of quantum-fluctuations in a 
parametric fashion. A natural choice for $D$ is the 
(non-interacting) bandwidth that according to
Fig. \ref{fig:lattice} (right) is given by $D\approx 6t$. 
Furthermore,
$\Delta^\text{DMRG}_\infty(\Uo{\approx} D)\sim t^2/D$ 
while $\Delta^\text{HF}_\infty(\Uo{\approx} D)\sim D$ 
so that we expect a large ratio 
$$
\left. \frac{\Delta^\text{HF}_\infty}{\Delta^\text{DMRG}_\infty}\right|_{\Uo\approx D} \sim \frac{D^2}{t^2}; 
$$
it takes values about $36$ broadly consistent with the factor 
$35\pm 10$ seen in Fig. \ref{f3a}(top).}

\begin{figure}[tb]
  \includegraphics[width=0.95\columnwidth]{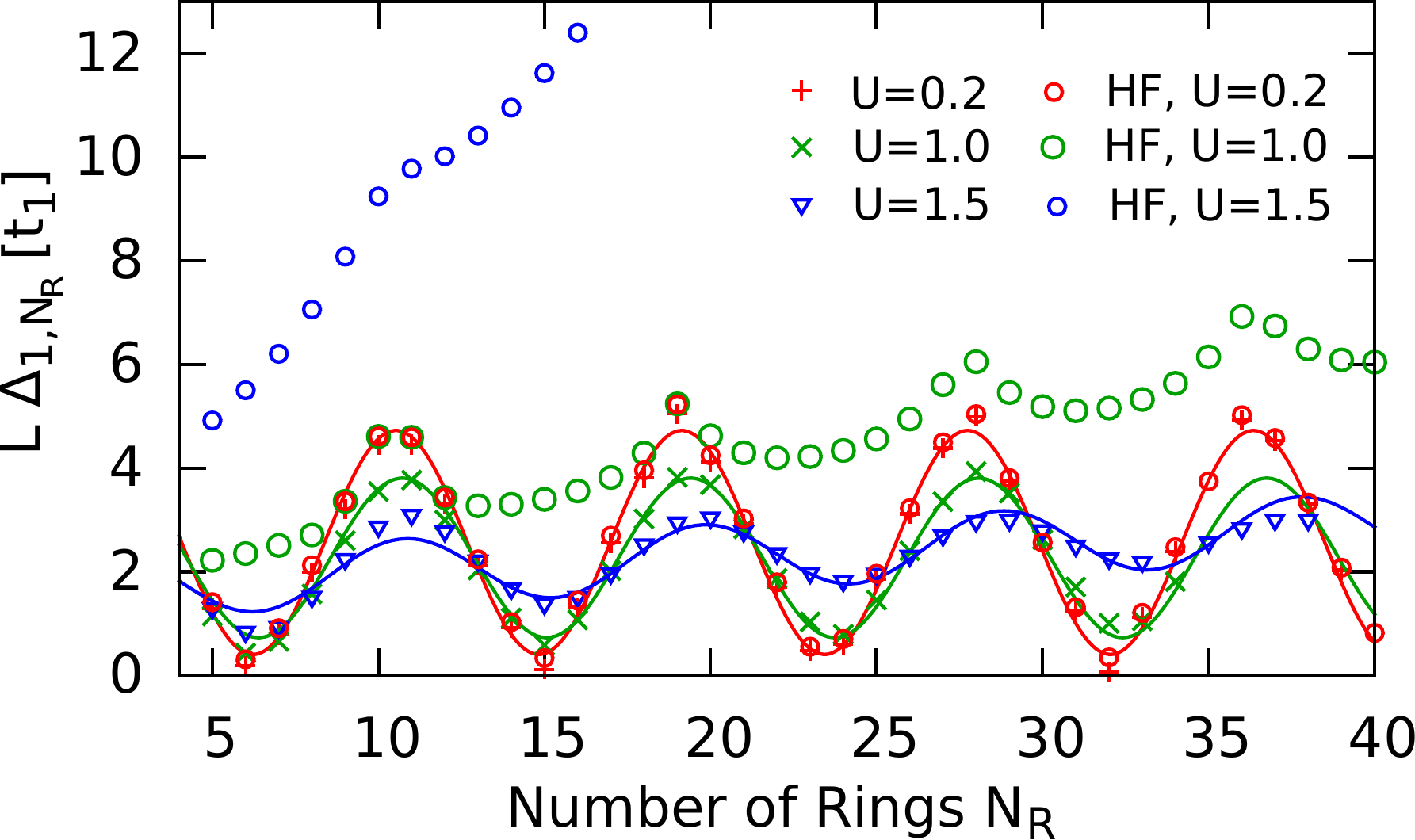}
  \caption{\label{f6}
  Evolution of the lowest lying excitation gap $\Delta_{1,\NR}$ for acene-ladders 
  with increasing number of rings $\NR$, $L=2\NR+1$, for varying interaction strengths
  $\Uo$ ($ +:0.2, \times: 1.0; \triangledown: 1.5$). 
  For comparison also the HF-result ($\circ$) is shown. 
  In contrast to the   strongly reduced \io{} on HF-level, 
  they survive for not too large $\Uo$ in the DMRG calculations. 
}
\end{figure}
Our final Fig.~\ref{f6} displays the most important result of our work. 
It confronts the mean-field data for the excitation gaps with the 
results obtained from the DMRG-calculations 
for finite-length wires. Specifically, the evolution of their DMRG-excitation gaps, 
$\Delta^\text{DMRG}_{1,\NR}$, is shown for increasing 
length of the wire $\NR$ at different $\Uo$.
It is clearly seen there, that the DMRG-traces exhibit pronounced \io{} 
even for those values of $U$ 
at which the residual oscillations in 
$\Delta^\text{HF}_{1,\NR}$ have almost faded away, already.
Very much in the spirit of the long-wire limit, we understand 
this result as a manifestation of quantum-fluctuations 
in finite-size wires. They destroy the tendency of mean-field 
theories to open up excitation gaps, thereby restoring the \io{} 
as a hallmark of the symmetry-unbroken phase.

\section{Conclusions}

We have studied the acene Hubbard model for a large range of on-site
repulsion $U$. The gap oscillations, that are characteristic of the 
weakly-correlated phase, survive as a periodic modulation also in the
strongly-correlated phase. The latter still persist for sizable interaction
strength, as quantum fluctuations significantly 
reduce the formation of the correlation gap.

With an eye on acene-molecules, we expect that  
our results have the following implications. 
We have performed model-calculations employing short-range 
interactions. They may apply to acenes in the presence of 
a screening environment, such as substrates or 
electrochemical solutions. With screened Coulomb-interactions, 
the effective value of $\Uo$ should be significantly smaller 
than the molecular bandwidth, implying that our 
weakly-correlated scenario should apply.
In contrast, for gas-phase molecules screening is very weak and 
the applicability of our model-calculation is not 
immediately guaranteed. 
This is the situation investigated 
in Refs. \onlinecite{wilhelm16,bettinger16,yang16}; we conclude 
with a brief overview. 

{\hi
Concerning Ref. \onlinecite{wilhelm16}, it remains to be 
seen if quantum-fluctuations can restore the spin-rotational symmetry
also in the presence of a long-range Coulomb interaction. 
Very recent quantum-chemistry calculations on the level of 
density-functional multi-reference theory (DFT/MRCI) for 
oligoacenes (up to $\NR{=}9$) and pp-RPA
(up to $\NR{=}12$)
may give an indication that this 
indeed could be the case.\cite{bettinger16,yang16} 

The authors of Ref. \onlinecite{bettinger16} 
confirm earlier findings\cite{mondal09} 
that best agreement between DFT/MRCI results and 
experimental IR-excitation energies is 
achieved for closed-shell reference states (RB3LYP) 
even though spin-unrestricted calculations may 
produce a lower total energy. It is thus suggested that 
quantum-fluctuations are indeed strong in long acene-wires 
so that single-determinant approximations 
of the ground-state are not representative.  
In agreement with this interpretation, these authors observe that 
the weight of the reference-state 
in the DFT/MRCI ground state is progressively decreasing 
with growing wire length, so that more and more Slater determinants
have to be included to accurately represent this state.

The authors of Ref. \onlinecite{yang16} come to a qualitatively 
similar conclusion. The pp-RPA method  is an interesting variant of
RPA-type approaches, since it incorporates vertex-corrections that 
are missed, e.g., in standard RPA bubble-summations. Thus, it allows to
include a certain amount of quantum fluctuations 
when treating unscreened wires (i.e. with
Coulomb-interactions) that would otherwise be too long for more accurate approaches. 
 A certain drawback of the method is 
is that it is uncontrolled: while it can singalize qualitatively 
{\em that} quantum fluctuations are likely to be important, there is no way of 
telling with confidence what they actually do in that case, quantitatively. 

On the qualitative level, the conclusions of
Ref. \onlinecite{bettinger16} and Ref. \onlinecite{yang16} are 
consistent. In particular, also the authors of 
\hi
\onlinecite{yang16} emphasize 
\hi
 that the multi-determinant character of the groundstate 
increases rapidly with growing wire length. It hinders the formation
of long-range magnetic order and implies that mean-field results 
(i. e. unrestricted open-shell calculations) are to be interpreted with a 
grain of salt. 

It remains to be seen, however, if restoring the spin-symmetry in the
ground state is enough to also restore the \io. Namely, even assuming 
that long-range and short-range models of molecular wires share the same phases, 
it is still unclear whether the effective value of $\Uo$
in the gas-phase calculations  is small or large as compared to
the (non-interacting) band-width. 

Interestingly, the authors of Ref. \onlinecite{yang16} propose an exponential decay 
of the lowest lying spin-triplet gap with increasing $\NR$ 
saturating near $\NR{=}11$ at roughly 0.13 eV without a trace of \io. 
If indeed correct, this finding could still be interpreted within the short-range 
model. It could be taken as a hint that the effective interaction $\Uo$ as it
appears in the study Ref. \onlinecite{yang16} should be considered large. 
However, the authors' data underlying their exponential fit (Table 1 in Ref. \onlinecite{yang16}) 
appears to be inconsistent with an exponential asymptotics. In fact, 
the data for the spin-restricted cases exhibits  even non-monotonic behaviour 
(upturn at large wire length), which is qualitatively consistent with 
the onset of \io{} as   predicted by us. 
\lo 

\begin{acknowledgments}
We express our gratitude to Holger Bettinger, Reinhold Fink, 
Steve Kivelson and Jan Wilhelm for enspiring discussions. 
Also, we thank Jan Wilhelm for sharing a manuscript 
\cite{wilhelm16} prior to publication. PS and RT are supported by
DFG-SFB 1170 and the ERC starting grant TOPOLECTRICS
(ERC-StG-336012). FE acknowledges support of the DFG under projects EV30/8-1 and EV30/11-1. 
Most of the calculations were performed on the compute cluster of the YIG group of Peter Orth.
\end{acknowledgments}

\appendix
\section{Non-interacting band-structure and an effective model}
\label{sec:appendixKorytar}
We consider a non-interacting ($U=0$) polyacene wire with the lattice
and single particle Hamiltonian of~\eqref{e1} explained
in Fig.~\ref{fig:lattice}. For this model, we derive analytic expressions for the
wave number of the nodal point $k_\mathrm D$, Fermi velocity $v_\mathrm F$, and the oscillation
period of gaps of finite chains.
The ladder-like lattice 
features a symmetry with a line group D$_{\infty\mathrm h}$. An important member of the group
is a mirror symmetry with respect to the interchange of stringers which we call parity.
The third-neighbor hoppings can be divided in two groups:
a hopping that connects atoms related by the parity and 
the two hoppings that connect different rungs (Fig.~\ref{fig:lattice}). 
We denote the former by $\tpp_\perp$ and the latter by $\tpp_\times$.
We label the 4 Bloch bands $\epsilon_{sb}(k)$ by the wave-number in units
of $a^{-1}$ (the inverse lattice spacing),
a parity $s=\pm 1$ and a remaining band index $b=\pm 1$. The explicit
expression for the band dispersions is given in~\eqref{disp}.
The bands are particle-hole symmetric only if $t'=0$, whence
 $\epsilon_{sb}(k) = -\epsilon_{-sb}(k)$.

The two bands with $b=-1$ give rise to a linear dispersion $bv_\mathrm F(k - k_\mathrm D)$ 
at the Fermi level if the condition $t\tpp_\perp >0$ is met, as we will show below. 
We derive analytic exressions for $k_\mathrm D$ and $v_\mathrm F$ under the condition $t'=0$.
The nodal (``Dirac'') point $k_\mathrm D$ for $t'=0$ is the solution of the equation
\begin{equation}
\cos(\pi - k_\mathrm D) = \frac{t^2 - \frac 12 t\,\tpp_\perp} {t^2 - \tpp_\perp \tpp_\times}. \label{eq:kD}
\end{equation}
In the limit of $(\pi - k_\mathrm D)$ small the nodal (Dirac) point and the Fermi
velocity are expressed by
\begin{align}
\label{Eq:kd}
\pi - k_\mathrm D &= \sqrt\frac{\tpp_\perp}{t} \left ( 1  - \frac{\tpp_\times}{t} \right )
\cdot \left [ 1 +  \mathcal O (\tpp/t)^2\right ] \\
v_\mathrm F &= 2\sqrt{t\, \tpp_\perp} \cdot \left[1 
- \frac{\tpp_\times}{t} - \frac 98 \frac{\tpp_\perp}{t} +  \mathcal O(\tpp/t)^2\right].
\end{align}
For oligoacenes within the tight-binding approximation parameterized as above, the
band crossing implies gap oscillations with the unit cell period of
\begin{equation}
\label{Eq:period}
 (1 - k_\mathrm D/\pi)^{-1} = \pi\sqrt {t\,  \tpp_\perp} \left( 1 + \frac{\tpp_\times}{t}
 + \mathcal O (\tpp/t)^2\right).
\end{equation}

In the equations~(\ref{Eq:kd}-\ref{Eq:period}) the symbol $\mathcal O(\tpp/t)^2$ denotes 
all terms that are of second order in $\tpp_\perp$ and $\tpp_\times$.
We see that the band crossing sets in with $\sqrt{\tpp_\perp}$, provided that the
signs of $t$ and $\tpp_\perp$ are equal.

The hopping $t'$ preserves the band crossing and weakly renormalizes $k_\mathrm D$ and $v_\mathrm F$. 
We remark that the low-energy behavior of the non-interacting polyacene model used in the
present paper can be mapped onto a simpler effective model with $t'=\tpp_\times = 0$ and $t$ and $\tpp_\perp$ 
non-zero, i.e. a ladder model with alternating rungs.
This model has been introduced originally by Kivelson and Chapman \cite{Kivelson}.

\section{Extracting the bulk gap from finite-size wires \label{a2}}
\label{app-gap}

Assuming that the gap opening only affects the band 
structure close to the Dirac point,
one would expect that the gaps are given by 
a minimal model of a type  
\begin{eqnarray}
\label{e3}
\Delta^\text{DMRG}(\NR) &{=}& 
\sqrt{\Delta(\NR)^2 + v^2_\text{F}\Delta k^2(\NR)} \\ 
\Delta k(\NR) &{=}& \frac{\pi}{2\NR+2} \cos(k_\text{D}\NR+\phi),
\end{eqnarray}
that follows the zone-folding argument of Ref. \onlinecite{korytar2014}. 
The sqrt term in Eq.~\eqref{e3} accounts for the level crossing
with parameters $\Delta(\NR)$, $v_\text{F}$, and  $\phi$; 
$\Delta(\NR)$ and $\phi$ are fit-parameters that accommodate additional 
finite-size effects.
\begin{figure}[htb]
  \centerline{\includegraphics[width=0.95\columnwidth]{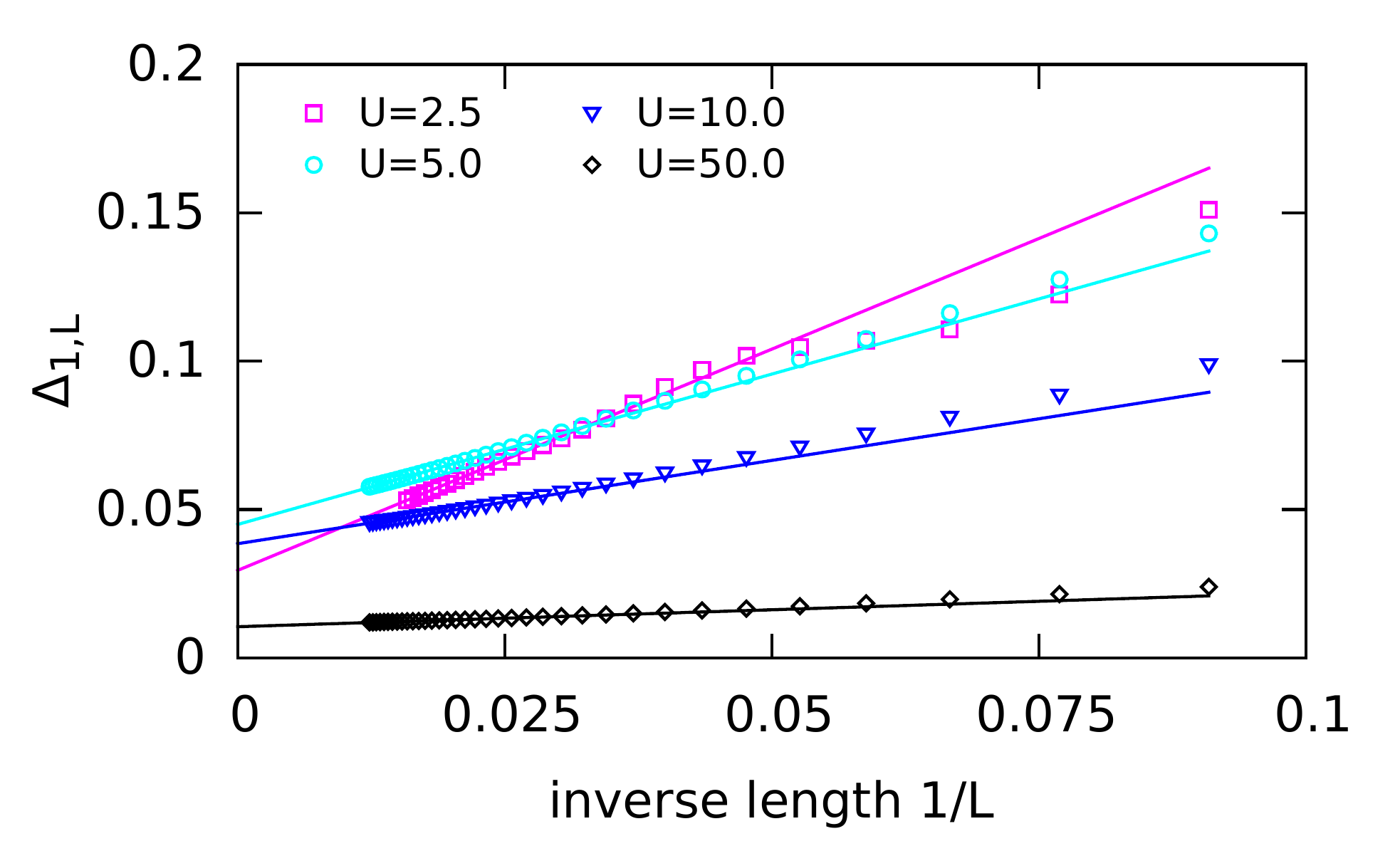}}
  \caption{ \label{fig:FitGap_LU}
  Extraction of the DMRG gap for large interaction values
  $\Uo$ ($\square$: $2.5$, $\circ$: $5.0$, $\triangledown$: $10.0$, $\diamond$: 50.0)
  via a linear fit of the finite site excitation gap $\Delta_{1,L}$ vs. the inverse system length
  $1/L$, $L=2\NR+1$.
  }
\end{figure}
Somewhat unexpectedly, our calculations indicate that the band-structure underlying Eq. \eqref{e3}  does not provide 
a faithful description of the true bandstructure of the infinite wire 
with the range of $\Uo$-values that we have been able to investigate.
At values $\Uo\sim 1$ the band-structure deviates significantly 
from the massive Dirac-shape and Eq. \eqref{e3} no longer applies.
(For instance we find a stronger damping of the oscillations as expected from Eq.~\eqref{e3} indicating an interaction
induced flattening of the band.)

In the limit of large interaction $\Uo$ the finite size oscillations are completely washed out and the 
thermodynamic limit of the interaction induced gap can be extracted from a linear fit in the inverse system size $1/L$, see Fig.~\ref{fig:FitGap_LU}.
However note, that for $\Uo=2.5$ an oscillatory part is already visible for not too large system sizes. 
Therefore we take an heuristic approach modeling the decay of the oscillations with a power law decay
\begin{align}
  \Delta_1(L) &=  \Delta_\infty +  a/L + b \cos\left( \kD L + \eta \right)/L^{1+\alpha} \,. \label{eq:FitGap} \,.
\end{align}
as shown in Fig.~\ref{fig:FitGap_Osc2}.
Here, $a/L$ describes the decay of the non-oscillatory part, while the cosine term describes a decay of the oscillation
faster than $1/L$, i.e. $\alpha >0$. $b$ gives the amplitude of the oscillations, $\kD$ gives periodicity
and $\eta$ allows for the shifting the phase of the oscillation. 
$\alpha$ is an heuristic exponent modelling the damping of the oscillations. In our fits we always find $\alpha >0$
and $\alpha$ approaches zero for $\Uo \rightarrow 0$.

\begin{figure}[hbt]
   \centerline{\includegraphics[width=0.95\columnwidth]{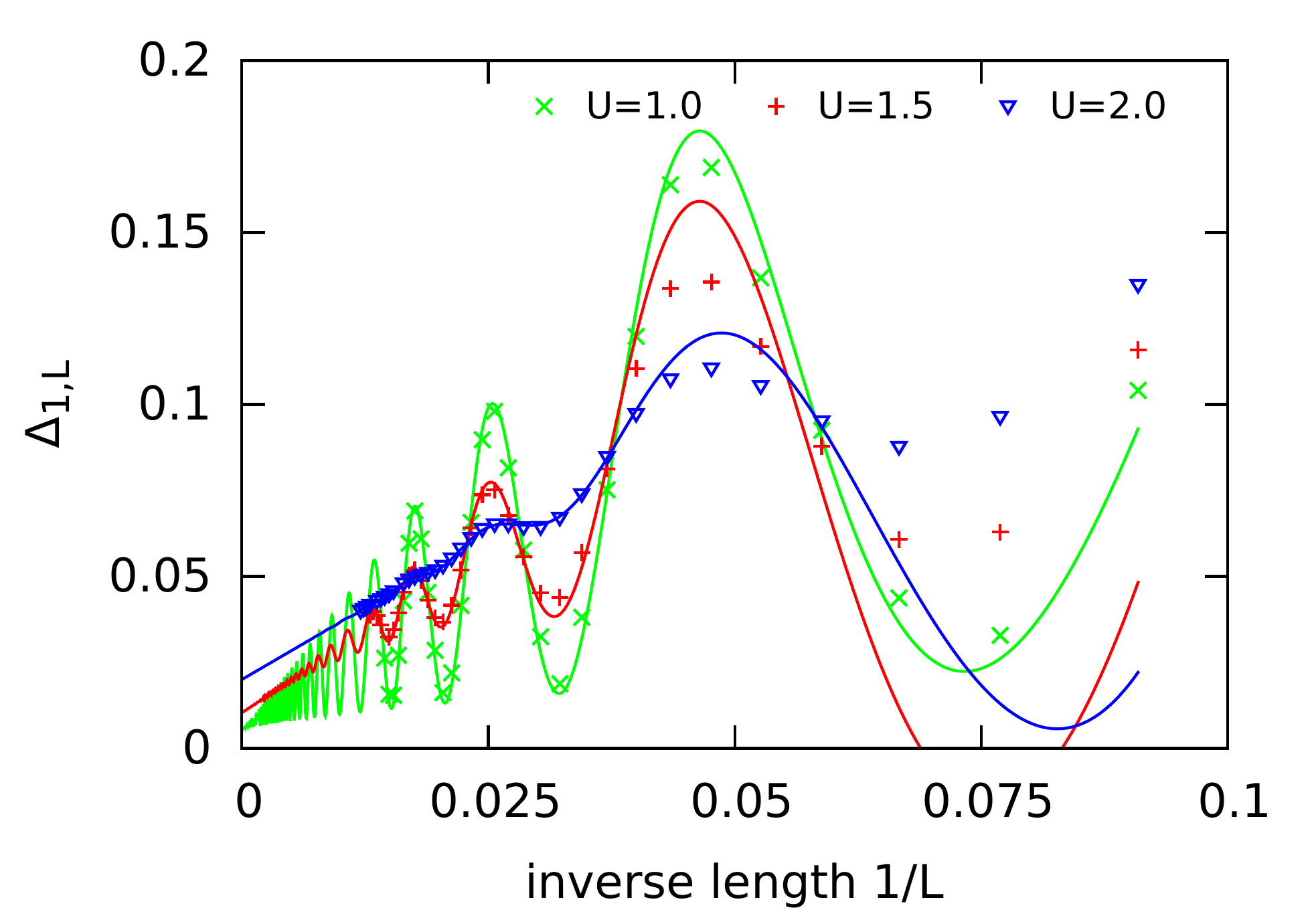}}
  \caption{\label{fig:FitGap_Osc2}
  Extraction of the DMRG  gap $\Delta_{1,\infty}$ for not too large interaction values
  $\Uo$ ($\times$: 1.0, $+$: 1.5, $\triangledown$: 2.0).
  The extrapolation is performed using Eq.~\eqref{eq:FitGap} for $\Delta_{1,L}$ vs.\ the inverse
  length $1/L$, $L=2\NR+1$, and displayed by lines, while the
  symbols correspond to the DMRG results.
}
\end{figure}

The gaps shown in Fig.~\ref{f3a} are obtained from the linear fits for $\Uo > 2.0$, see Fig.~\ref{fig:FitGap_LU},
and from fits of Eq.~\eqref{eq:FitGap} for $\Uo\le 2.0$.
We mention that the fit for $\Uo=1.0$ is still sensitive to
the fit range, and the result should be taken with care. 
For this reason we refrained from including results 
for $\Uo<1.0$. In addition, in that regime
it is  not sufficient to fit a single cosine, as the results are getting close to the 
saw tooth like behavior as in the non-interacting case.
%
\begin{figure}[hbt]
   \centerline{\includegraphics[width=0.95\columnwidth]{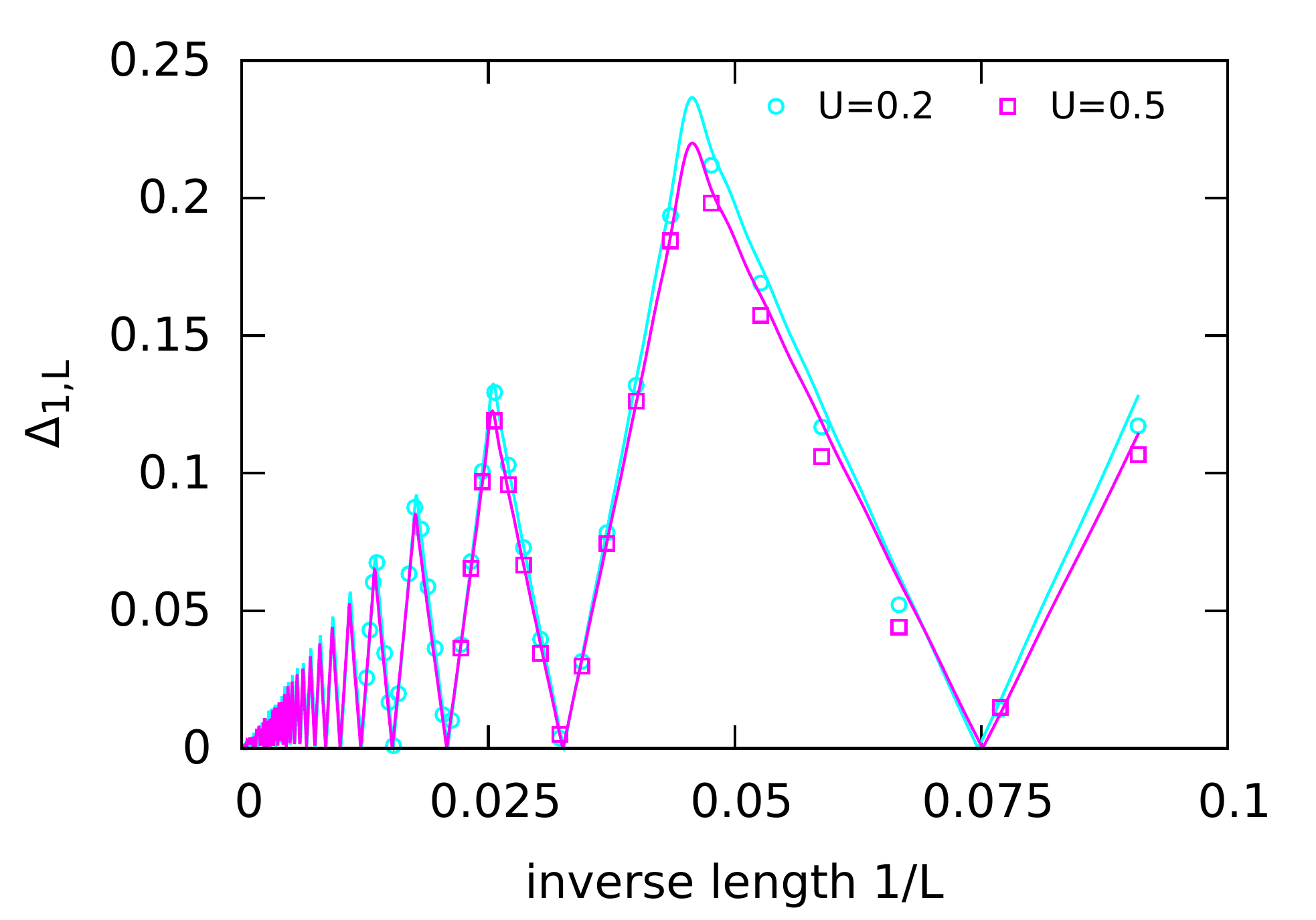}}
  \caption{\label{fig:FitGap_Osc1}
  Extraction of the DMRG gap  $\Delta_{1,\infty}$ for small interaction values
  from  $\Delta_{1,L}$ vs. the inverse length $1/L$, $L=2\NR+1$.
  For not too large $\Uo$ ($\times$: 0.2, $\circ$: 0.5).
  The fits correspond to Eq.~\eqref{e3} where the cosine is replaced by a saw tooth Eq.~\eqref{eq:RoundedSawTooth}
}
\end{figure}
%
In order to fit the results for small $\Uo$ we introduce a (truncated) saw tooth function
\begin{equation}
 g(x) = \frac{\sum_{\ell=1,3,5,\cdots, }^n  \cos( \ell x  )/\ell^2}{ \sum_{\ell=1,3,5,\cdots}^n \ell^{-2}} \label{eq:RoundedSawTooth}
\end{equation}
and use it instead of the cosine in Eq.~\eqref{e3}. Here we imposed a limit of n=17. 
The results for $\Uo=0.2$, $0.5$ are displayed in Fig.~\ref{fig:FitGap_Osc1}.
While the fits do capture the oscillation quiet well, the finite size oscillations are orders of magnitudes larger than
the extracted gaps. In return we do not trust the extracted values and
did not include them in Fig.~\ref{f3a}.

\section{Methodological details}
\label{sec:HFNumericsDetails}
\subsection{Numerical Hartree-Fock}

Due to  SU(2) symmetry combined with a local on-site interaction, we can restrict our HF equations to the Hartree terms
\begin{equation} \label{eq:HFnum}
 \HH^{\mathrm{HF}} = \hat{H}^{U=0} \,+\, U \sum_{x,\sigma} \left( \bar{n}_{\sigma,x} - 0.5\right) \hat{n}_{-\sigma,x}  \,,
\end{equation}
where $\hat{H}^{U=0}$ denotes the non-interacting system according to Eq.~\eqref{e1},
$\sigma$ denotes up and down spins, $-\sigma$ the opposite spin to $\sigma$,
and $\bar{n}_{\sigma,x}$ the expectation value of the local density operator $\hat{n}_{\sigma,x}$ at site $x$ with spin $\sigma$.
We have checked that by breaking the remaining $S_{\text{tot}}^z$ conservation, therefore allowing for Fock terms of the form
$\hat{c}^\dagger_{\sigma,x} \hat{c}^{}_{-\sigma,x}$,
we only find solutions where the spin quantization axis is  rotated with respect to the Hartree solution.
Note that in this case the solution depends on  the initialization of the self consistency loop.

Since up and down spins do not mix in~\eqref{eq:HFnum}, we can represent the up and down spin sectors
by independent matrices. 
We start the HF calculations by first diagonalizing the non-interacting system in order to obtain the ground
state energy $E_{0}$ as a reference. Note that in this case we obtain a homogeneous solution $\bar{n}_{\sigma,x} = 0.5$.
We then explicitly introduce a staggered magnetic order by setting 
$ \bar{n}_{\sigma,x}  = 0.5 \pm 0.01 \sigma$, where we take '$+$' on one sublattice of the bipartite system
and '$-$' for the other sublattice (Fig.~\ref{fig:lattice}). 
With these initial values we diagonalize $\HH^{\mathrm{HF}}$ and calculate the locate densities $\bar{n}_{\sigma,x}$
which we then iteratively insert into~\eqref{eq:HFnum}. We perform at least five iterations and continue until
the ground state energy $E^{\mathrm{HF}}$ of Hamiltonian  $\HH^{\mathrm{HF}}$ changes by less then $10^{-10}t$. 
In order to avoid getting stuck in an oscillation between two solutions,
we damp the HF self consistency loop by taking an average of 0.7 times
the new density and 0.3 times the densities of the preceding iteration.

Finally, we check whether the ground state energy  $E^{\mathrm{HF}} $ including the double counting corrections
\begin{equation}
  E^{\mathrm{HF}} = E^{\mathrm{SC}}  \;-\; \frac{U M}{4} \;+\; U \sum_j \bar{n}_{\sigma,x} \bar{n}_{-\sigma,x} 
\end{equation}
is smaller than the ground state energy $E_{0}$ of the non-interacting system.
Indeed we always find $ E^{\mathrm{HF}} \le E_{0}$ 
The spin-averaged local density $\bar{n}_{\uparrow,j} + \bar{n}_{\downarrow,j}$ of our HF solution is homogenous
and the total staggered magnetisation $M^{\mathrm{HF}}$ is given by the difference between the 
magnetization of the two sub lattices
\begin{align}
  m_j  &= \frac{1}{2} \left( \bar{n}_{\uparrow,j} - \bar{n}_{\downarrow,j} \right), \\
  M^{\mathrm{HF}} &= \sum_{j\in A} m_j \;-\;  \sum_{j\in B} m_j  \;.
\end{align}

\subsection{Hartree-Fock: commensurate case}
\label{sec:HFCommensurateSystem}
\begin{figure}[t]
  \includegraphics[width=0.95\columnwidth]{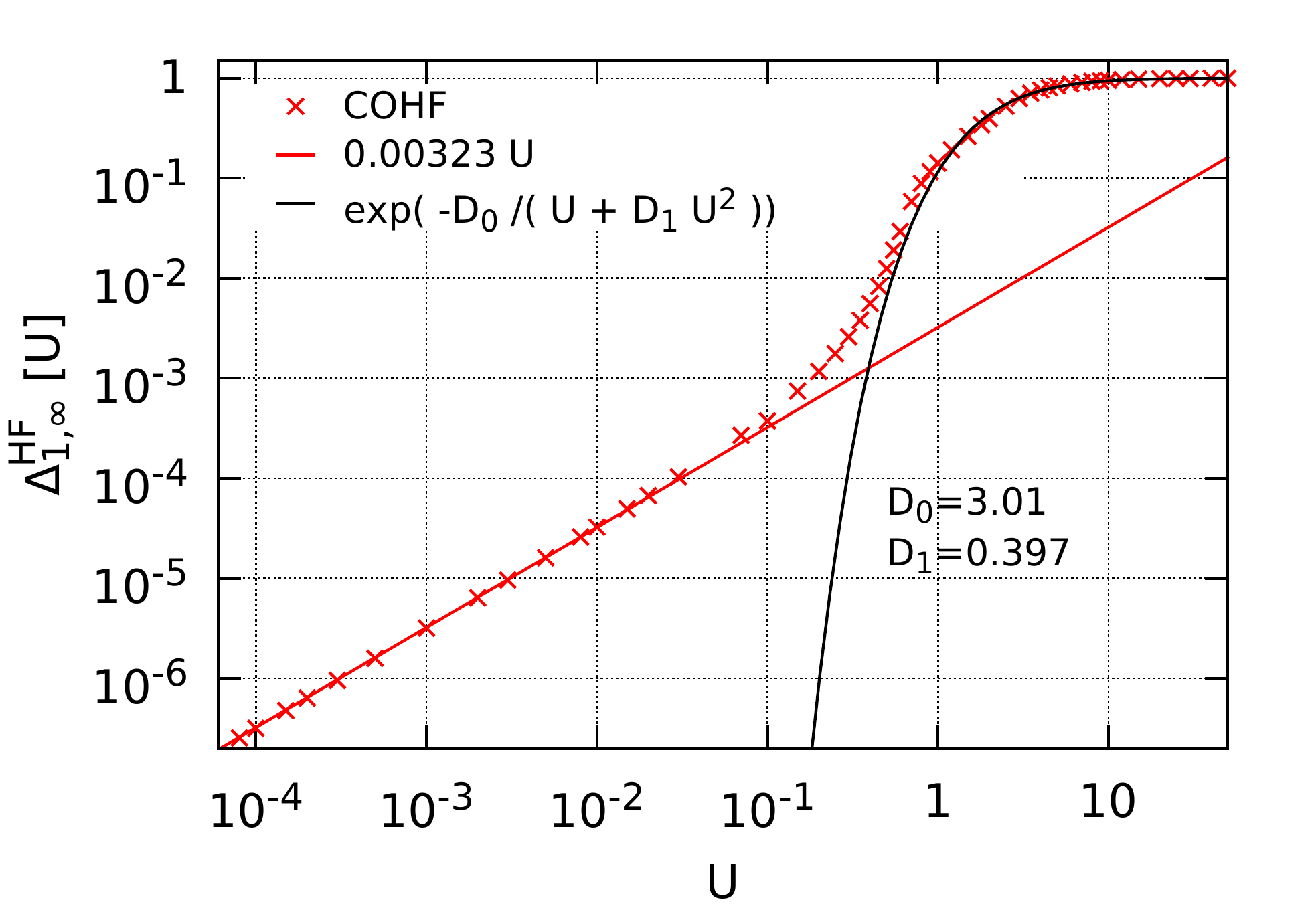}
  \caption{\label{fig:HFgapCO} 
   The HF gap $\Delta^{\mathrm{HF}}_{1,\infty}$ as shown in Fig.~\ref{f3a}, left, in the case of commensurate oscillations $\tpp=0.13$ where
   the periodicity is given by 20 rings. The gap is obtained by HF calculations for $\NR=20, 40, 60, \cdots, 200$.
 }
\end{figure}

The main problem of obtaining the $U\ll 1$ HF solutions in Fig.~\ref{fig:GapsIO} roots in the finite size oscillations
being incommensurable. Therefore we need very large system sizes in
order to start with an $U=0$ finite size gap $\Delta_0$ being smaller
than the interaction induced gap in the HF calculations.
This extraction of the HF gap gets simplified if we switch to the case of commensurate oscillations by choosing a fine-tuned $t^{''}$
from our solution Eq.~\eqref{eq:kD}. In Fig.~\ref{fig:HFgapCO} we display the results in the case of a periodicity of twenty rings.
We obtained the corresponding $t^{''}=0.13$ by solving Eq.~\eqref{eq:kD} numerically.%
\footnote{For reference, the precise value used in the numerics is $\tpp=0.130064666287555$.}
Since we now always hit a minimum of the gap oscillations precisely, i.e. a zero in the non-interacting case, we can extract
the HF gap without difficulties. As expected, the HF gap indeed opens
$\Delta^{\text{HF}}_{1,\NR}\sim \Uo^2$. In addition, we see the same
exponential behavior for intermediate to large $\Uo$
as displayed in Fig.~\ref{f3a}.

\subsection{DMRG}
\label{sec:DMRGDetails}
Within our DMRG  calculations we keep up to 8000 states per 
block; the dimension of the target space grows up
to $4.6 \cdot 10^7$, the discarded entropies are 
considerably below $10^{-4}$  for $U>1.0$,
while for $U\le1.0$ the discarded entropies reach $4.2  \cdot  10^{-2}$
for systems with $\NR >20$
in each DMRG step, typically significantly smaller. We employ at least 7 
finite lattice sweeps with an optimized infinite lattice warm up for the smallest system sizes.
In order to reach large system sizes we iteratively restart a $\NR$ simulation in order to investigate 
a $\NR +1$ ring. Note that for this we only have to deactivate the wave function prediction
in the initial step of a restart.
In order to initialize the simulations we apply a sliding environment block approach,
that we already employed successfully to fractional quantum hall systems.\cite{Hu:PLA12,Johri:PRB14,Johri:NJP16}
Instead of using a reflection or a growing right block during the  DMRG infinite lattice warm up sweeps,
we employ a right block of up to 5 sites, which we can always build
exactly in such a way that we always have $4 \NR$ sites for PBCs
and $4 \NR + 2$ sites for HWBCs. With this initial warm up, we avoid running into
excited states in the initial sweep.
%
\bibliographystyle{unsrt}
\bibliography{references}
\end{document}